\documentclass{aa}

\usepackage{graphicx}
\usepackage{txfonts}

\usepackage{xcolor}
\usepackage{xspace}
\usepackage[colorlinks,urlcolor={blue!80!black},citecolor={blue!60!black},linkcolor={red!60!black}]{hyperref}

\newcommand{\frbpoppy}{\texttt{frbpoppy}\xspace}
\def\pop#1{\texttt{#1}\xspace}
\def\survey#1{\texttt{#1}\xspace}
\def\est#1{\textcolor{gray}{#1}\xspace}
\newcommand{\paperone}{2019A&A...632A.125G}
\newcommand{\package}[2]{\texttt{#1} \citep{#2}}

\makeatletter
\renewcommand*\aa@pageof{, page \thepage{} of \pageref*{LastPage}}
\makeatother
\begin{document}
\title{Synthesizing the repeating FRB population using \frbpoppy}

\author{D.W. Gardenier 
 \inst{1, 2}\fnmsep\thanks{\email{gardenier@astron.nl}}
 \and
 L. Connor\inst{2}
 \and
 J. van Leeuwen\inst{1, 2} 
 \and
 L. C. Oostrum\inst{1, 2}
 \and
 E. Petroff\inst{2}
}
\institute{ASTRON, the Netherlands Institute for Radio Astronomy, Oude Hoogeveensedijk 4, 7991 PD, Dwingeloo, The Netherlands
 \and
 Anton Pannekoek Institute for Astronomy, University of Amsterdam, Science Park 904, 1098 XH Amsterdam, The Netherlands
}

\date{Received October 9, 2020} 

\abstract{The observed Fast Radio Burst (FRB) population can be divided into one-off and repeating FRB sources.
 Either this division is a true dichotomy of the underlying sources, or selection effects and low activity prohibit us from observing repeat pulses from all constituents making up the FRB source population.
 We attempt to break this degeneracy through FRB population synthesis.
 With that aim we extend \texttt{frbpoppy}, which earlier only handled one-off FRBs, to also simulate repeaters.
 We next model the Canadian Hydrogen Intensity Mapping  Experiment FRB survey (CHIME/FRB).
 Using this implementation, we investigate the impact of luminosity functions on the observed dispersion measure (DM) and distance distributions of both repeating and one-off FRBs.
 We show that for a single, intrinsically repeating source population with a steep luminosity function,
 selection effects should shape the DM distributions of one-off and repeating FRB sources differently.
 This difference is not yet observed.
 We next show how the repeater fraction over time can help in determining the repetition rate of an intrinsic source population.
 We simulate this fraction for CHIME/FRB, and show that a source population comprised solely of repeating FRBs can describe CHIME/FRB observations with the use of a flat luminosity function.
 From the outcome of these two methods we thus
 conclude that all FRBs originate from a single and mostly uniform population of varying repeaters.
 Within this population, the luminosity function cannot be steep,
 and there must be minor differences in physical or behaviour parameters that correlate with repeat rate.
}
\maketitle

\section{Introduction}
Fast Radio Bursts (FRBs) are millisecond duration pulses detected at radio frequencies \citep{cordes2019review,petroff2019review}.
At least over 21 FRB sources have been observed to repeat (repeaters), with 127 FRB sources not observed to have repeated (one-offs) \citep{frbnewsletter11}.
Originally, FRBs were by-catch of pulsar surveys, but since about 2018 dedicated FRB surveys have begun operation.
The main three observatories searching for FRBs include the Canadian Hydrogen Intensity Mapping Experiment (CHIME; \citealt{chimeoverview}), the Australian Square Kilometre Array Pathfinder (ASKAP; \citealt{craft, askap}) and Apertif on Westerbork \citep{arts, lk+21}.
While initially each new FRB detection was considered newsworthy \citep[e.g.][]{masui2015}, the rise in FRB detections through these FRB surveys has ushered in the dawn of FRB population studies \citep{macquart2019}.
Initial population studies had few FRBs with which to work \citep{thornton2013,macquart2015}, however subsequent studies investigating detection biases \citep{macquart2018paperone}, rate distributions \citep{james2020} or spectral properties \citep{macquart2019} were able to utilise a larger sample of FRBs.\\

The detection of a repeating FRB source in 2016 \citep{r1} already prompted the question of whether all FRB sources repeat.
Do both apparent types of FRBs emerge from the same intrinsic source population?
Despite extensive observational campaigns \citep[e.g.][]{petroff2015, 2018Natur.562..386S}, no conclusive evidence has emerged either way.
Theoretical studies of possible FRB source mechanisms provide no conclusive answer either. Models such as neutron star -
white dwarf accretion \citep{gu2016}, supergiant pulses \citep{cordes2016supergiant}, blast waves from magnetars
\citep{metzger2019} or emission within neutron star magnetospheres \citep{2020arXiv200505093L} can produce both
repeaters and one-offs. \\

One possible approach to probing the intrinsic source class, is population synthesis.
In studies of pulsars \citep{taylor77}, gamma ray bursts \citep{GRBs}, and stellar evolution \citep{2018arXiv180806883I}, population synthesis has proved to be a powerful tool.
To this end, we previously implemented an open-source FRB POPulation sythesis package in PYthon \citep[\frbpoppy;][]{\paperone}.
This first version was capable of modelling one-off FRBs and could successfully reproduce the observed one-off FRB populations as seen by the High Time Resolution Universe (HTRU) survey and by ASKAP.
In this paper we present an updated version of \frbpoppy capable of modelling repeating FRB sources.
We use it to probe the intrinsic FRB source population in multiple ways.
These methods could allow the field to determine the nature of the FRB population.\\

Prior population synthesis efforts by \citet{caleb2019} simulated repeaters with a variety of wait time distributions to determine expected detection rates and constraints on the slope of the intrinsic energy distribution.
\frbpoppy takes a different approach, with increased focus on survey modelling to replicate a wide range of selection effects.
Additionally, \frbpoppy has been designed from the ground up to be an open source, modular Python package for easy use by the community \citep{\paperone}.
Simulations of the repeating FRB population are increasingly being used to probe various aspects of the intrinsic FRB population \citep[e.g.][]{ai2020}, but often lack the modelling of the full range selection effects present in the observed FRB population, which are of essential importance.\\

In this paper we use population synthesis to show several methods by which the intrinsic FRB source population can be constrained.
We start by detailing our approach to synthesizing a repeating FRB population before providing our results and interpretation thereof in the second half.
As such, we present an implementation of repeating sources in Sect.~\ref{sect:methods}, in terms of both generating and surveying repeating sources.
We then show several ways by which population synthesis can identify selection effects in the observed FRB populations, through which the intrinsic FRB source population can be probed in Sect.~\ref{sect:results}.
We subsequently summarise our thoughts in Sect.~\ref{sect:conclusion}.
The paper ends with Appendix~\ref{sec:appendix}, containing information relevant to our methods.

\section{Simulating a repeater population}
\label{sect:methods}
Population synthesis is a method by which properties of an underlying, real source population are derived by simulating virtual populations \citep[see e.g.][]{taylor77}.
To this end, we presented \frbpoppy in \citet{\paperone}: a code base capable of modelling one-off FRBs and thus constraining properties of the intrinsic FRB source population.
Additional constraints on the FRB source population can however be found by looking to repeating FRB sources \citep[see e.g.][]{fonseca2020}.
We aim to take advantage of repeater observations by incorporating repeating sources into \frbpoppy.
These features can be found in the \texttt{v2} release of \frbpoppy, accessible on Github\footnote{https://github.com/davidgardenier/frbpoppy}.\\

Shifting from one-off FRB sources to repeating sources requires additional \frbpoppy functionality in three major areas: in simulating burst times, in generating properties and in surveying populations.
This functionality is described in the following sections.
In describing such population synthesis methods,
the term `observed population' can be confusing as there are both real and simulated `observed populations'.
Often the interpretation can be gained from the context, but where this is lacking, we ensure the terms real or simulated are added.
Furthermore, the term FRB originally referred to both the  burst and the source.
For repeaters, these are different concepts.
Throughout this paper we use `burst' to refer to an individual flash of light and `source' to refer to an origin of these bursts.
We thus use the term Fast Radio Burst to refer to a single burst.
To distinguish software input from other connotations we use a recognisable typeface, e.g. \texttt{chime-frb} as an argument versus CHIME/FRB the survey.\\

\subsection{Generating burst times}
Where simulations of one-off FRBs can be relatively static, repeaters require the simulation of repetition.
In \frbpoppy we generate a series of burst time stamps per FRB source.
A variety of distributions can be used to generate these time stamps, including:

\paragraph{\texttt{single}} To simulate one-off sources, the \texttt{single} option generates a single time interval per source within a given time frame.
\begin{equation}
 t_{\text{interval}} \in U(0, n_{\text{days}})
\end{equation}
Here time intervals ($t_{\text{interval}}$) are drawn from a uniform distribution $U$ in the range zero to the chosen maximum number of days $n_{\text{days}}$.

\paragraph{\texttt{regular}} To replicate  pulsars  \citep{hewish1968}, and for testing purposes, we allow for perfectly regular time intervals.
\begin{equation}
 t_{\text{interval}} = \frac{1}{r}\ k
\end{equation}
with rate $r$ and integer $k$, an iterator such that the maximum value of $t_{\text{interval}}$ remains smaller than the
maximum timescale ($n_{\text{days}}$).
The rate $r$ can vary per source.

\paragraph{\texttt{poisson}}
We can draw bursts from a Poissonian distribution, similar to the giant-pulse behaviour in pulsars \citep{1995ApJ...453..433L}.
We use the inverse cumulative distribution function (CDF) of an exponential function.
In this case, the probability density function (PDF) can be described as
\begin{equation}
 P(x) = r e^{-rx}
\end{equation}
for the rate $r$ when $r\ge 0$. From this the inverse CDF can be derived:
\begin{equation}
 t_{\text{interval}} = -\frac{\text{ln}(u)}{r}
\end{equation}
with rate $r$ and $u \in U(0, 1)$ where $U$ represents a uniform distribution.
To ensure enough bursts are simulated per source, bursts are drawn per FRB source until the cumulative time interval would result in a burst beyond the requested maximum timescale ($n_{\text{days}}$).
This last time interval is subsequently masked.
To simulate a variety of FRB sources, the rate $r$ can be chosen to vary per source.

\paragraph{\texttt{clustered}}
As FRB121102 follows a distinctly non-Poissonian burst rate \citep{oppermann2018},
\frbpoppy can  simulate such clustered bursts.
Time intervals are now drawn from the inverse CDF of the Weibull distribution.
The PDF of a Weibull distribution can be described as
\begin{equation}
 P(t_{\text{interval}}) = \frac{k}{\lambda}\left(\frac{t_{\text{interval}}}{\lambda}\right)^{k-1}e^{-(t_{\text{interval}}/\lambda)^{k}}
\end{equation}
with scale parameter
\begin{equation}
  \lambda = \frac{1}{r\Gamma(1 + 1/k)}
\end{equation}
with gamma function $\Gamma$, shape parameter $k$ and the rate parameter $r$ for $t_{\text{interval}} \geq 0$, from which the inverse CDF can be derived:
\begin{equation}
 t_{\text{interval}} = \frac{1}{r\Gamma(1 + 1/k)}\Big(-\text{ln}(u)\Big)^{1/k}
\end{equation}
with the rate parameter $r$, gamma function $\Gamma$, shape parameter $k$ and $u \in U(0,1)$ with uniform distribution $U$. Just as with the \texttt{poisson} option, bursts are iteratively generated up to the maximum timescale ($n_{\text{days}}$).\\

\paragraph{\texttt{cyclic}} With several repeaters showing quasi-periodic activity \citep[see][]{2020Natur.587...59B, 2020MNRAS.495.3551R, 2020MNRAS.500..448C}, the \texttt{cyclic} option allows \frbpoppy to model bursts emerging during an active window.
For simplicity, we model the arrival times of bursts during the active window as a uniform distribution:
\begin{equation}
    t_{\text{arrival}} \in U(0, n_{\text{active}})
\end{equation}
with $n_{\text{active}}$ the number of active days per activity cycle of the source. The number of generated bursts is given by
the product of the number of bursts per active period and the number of activity cycles within the maximum timescale ($n_{\text{days}}$):
\begin{equation}
    n_{\text{burst}} = r\,n_{\text{active}}\,\frac{n_{\text{days}}}{P}
\end{equation}
with burst rate $r$ and an activity cycle of $P$ days. For each next activity cycle, a whole period is added to the generated burst
 time to convert them into time stamps between 0 and $n_{\text{days}}$.\\

To generate time stamps from the time intervals given in some of these distributions, we take cumulative time intervals per source:
\begin{equation}
 t_{\text{stamp}} = \sum_{n=0}^{N} t_{\text{interval},\ n}
\end{equation}
with time stamp $t_{\text{stamp}}$, the $N$th burst of a source, and $t_{\text{interval}}$ the time interval since the previous burst.
All time stamps are subsequently scaled using
\begin{equation}
 t_{\text{measured}} = t_{\text{stamp}}(1+z)
\end{equation}
to obtain the measured time stamp $t_{\text{measured}}$ from the intrinsic time stamp $t_{\text{stamp}}$ and $z$ the redshift of the source. All bursts with measured time stamps falling outside of the requested time frame $n_{\text{days}}$ are masked.\\

The number of generated bursts per source is used to determine the number of values required in generating subsequent burst parameters.

\subsection{Generating repetition properties}
The repeat bursts of an individual FRB source can have quite different properties \citep[e.g.][]{r1,r2,gourdji2019,oostrum2020repeating}.
The observed burst luminosities for a single source may, for instance, fall in a narrower range than the luminosities spanned by the full repeater population.
Similarly, some repeating sources may repeat more often than other sources \citep{fonseca2020}.
These cases show the need to expand \frbpoppy capabilities beyond the  single distributions used in \texttt{v1}.
We need an overarching population distribution that provides input to source distributions. \\

For parameters unrelated to the location of a source (e.g. pulse width), we have adapted \frbpoppy to allow input parameters to be drawn from an overarching distribution per source.
The mean of an intrinsic Gaussian pulse width distribution per source can for instance be drawn from a log-normal population distribution.
Additional settings provide the opportunity to keep a constant parameter value per source (e.g. to simulate standard candles), or to draw all values from the same overarching distribution, irrespective of source.
To adopt these settings in \frbpoppy we use the argument `per source', being either the `same' per source (using a constant value per source), or `different' (drawing a new value for each burst of a source).

\subsection{Surveying repeater populations}
\label{sec:methods:surveying}
For modeling the observations, repeating sources pose a greater  challenge than one-off sources.
One-off bursts have an equal chance of falling anywhere within a beam pattern \citep[see][]{\paperone}.
This no longer holds when considering repeating sources - here, the locations in the beam pattern of multiple bursts from a single source are correlated.
Especially for e.g. regularly emitting repeaters, the exact beam shape then becomes important for recognizing an FRB as a repeater.
Accounting for this behaviour requires modelling and tracking the location of sources within a beam pattern over time.
Depending on the beam pattern, mount type and location of a survey, celestial objects track different paths throughout the beam.
For CHIME, a transit telescope \citep{chimeoverview}, the location can be described relative to the centre of the beam
pattern and the North-South and East-West axes, but this does not necessarily hold for other mount types.
In appendix~\ref{sec:appendix} we present the tracking implementation in \frbpoppy for a variety of mount types.
These effects have been modelled in \frbpoppy to ensure an accurate portrayal of any resulting detection rates.\\

Where with one-offs a one dimensional beam pattern suffices, a realistic simulation of the repeating population,
as detected with large, asymmetric beams such as used by CHIME, requires two dimensional beam patterns.
For surveying repeating populations we use the formulas given in \citet{\paperone} to simulate Gaussian and Airy beam patterns as 2D matrices, using a Field-of-View (FoV) parameter to scale beam patterns relative to the survey.
The Apertif and HTRU beam patterns available in \frbpoppy can be scaled in a similar manner.
Empirically mapping of the CHIME beam patterns is as of yet ongoing \citep[see][]{berger2016}, so to enable \frbpoppy to conduct a \texttt{chime-frb} survey, we model our own CHIME-like beam pattern.
To simulate this beam pattern, we convolve an Airy disk pattern orthogonal to a cosine function subtending a $80 \degr$x$180\degr$ area of the sky.\\

\begin{figure}
 \centering
 \includegraphics[width=\hsize]{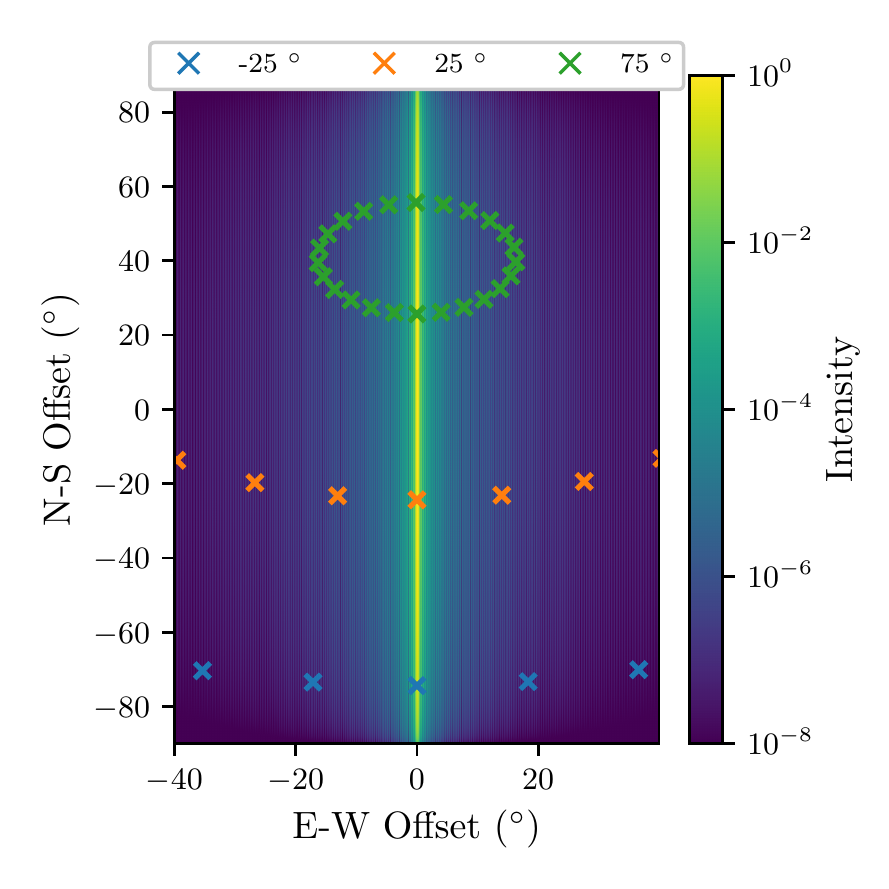}
 \caption{Simulated beam pattern for the CHIME/FRB survey showing the transit of sources at various declinations as orthogonal offset along North-South (N-S) and East-West (E-W) axes with respect to the centre of the beam pattern. All pointings are separated by an hour.}
 \label{fig:obj_beam}
\end{figure}

In Fig.~\ref{fig:obj_beam}, we show this CHIME-like beam pattern, with simulated observed tracks of several regular
emitters at various declinations.
The axes in Fig.~\ref{fig:obj_beam}, N-S/E-W offset, refer to the relative offset along the North-South and East-West axes with respect to the center of the beam pattern.
As expected, objects close to the North Pole are permanently visible.
Objects at low declinations transit the beam.
As all objects were emitting at the same cadence, the resulting spacing shows the transit speed.\\

Sets of survey parameters in \frbpoppy allow it  to model a range of current and future surveys \citep{\paperone}.
Additional parameters such as mount type and telescope location are required for surveying repeater populations.
These have been included in \texttt{v2}.
Table~\ref{tab:surveys} lists the main survey parameters adopted in the current paper.\\

\begin{table}
 \caption{An overview of the survey parameters adopted within this paper for a \survey{perfect} and a \survey{chime-frb} survey.
  Parameters include survey degradation factor $\beta$, telescope gain $G$, pointing time $t_{\textrm{point}}$, sampling time $t_{\textrm{samp}}$, receiver temperature $T_{\textrm{rec}}$, central frequency $\nu_{\textrm{c}}$, bandwidth BW, channel bandwidth BW$_{\textrm{ch}}$, number of polarisations $n_{\textrm{pol}}$, field-of-view FoV, minimum signal-to-noise ratio S/N, observatory latitude $\phi$, observatory longitude $\lambda$, mount type, and then the minimum to maximum right ascension $\alpha$, declination $\delta$, Galactic longitude $l$, and Galactic latitude $b$.
  \survey{chime-frb} survey parameters have been taken from the CHIME system overview paper \citep{chimeoverview}, with greyed out values indicating an estimated value or an average between given values. All \survey{perfect} survey parameters are necessarily self-devised.
 }
 \label{tab:surveys}
 \centering
 \begin{tabular}{c c c c}
  \hline\hline                                                            \\[-9px]
  Parameter            & Units   & \survey{perfect} & \survey{chime-frb}      \\
  \hline                                                                  \\[-9px]
  $\beta$              &         & {1.2}            & \est{1.2}           \\
  G                    & K/Jy    & {10$^5$}         & \est{1.4}           \\
  $t_{\textrm{point}}$ & s       & {86400}          & \est{360}           \\
  $t_{\textrm{samp}}$  & ms      & {0.001}          & 1                   \\
  $T_{\textrm{rec}}$   & K       & {0.01}           & 50                  \\
  $\nu_{\textrm{c}}$   & MHz     & {1000}           & 600                 \\
  BW                   & MHz     & {800}            & 400                 \\
  BW$_{\textrm{ch}}$   & MHz     & {0.001}          & 0.390625            \\
  $n_{\textrm{pol}}$   &         & {2}              & 2                   \\
  FoV                  & deg$^2$ & {41253}          & \est{164.15}        \\
  S/N                  &         & {10$^{-16}$}     & \est{10}            \\
  $\phi$               & \degr   & {0}              & 49.3208             \\
  $\lambda$            & \degr   & {0}              & -119.624            \\
  Mount                &         & {azimuthal}      & transit             \\
  $\alpha$             & \degr   & {0 -- 360}       & 0 -- 360            \\
  $\delta$             & \degr   & {-90 -- 90}      & \est{-40.679} -- 90 \\
  $l$                  & \degr   & {-180 -- 180}    & -180 -- 180         \\
  $b$                  & \degr   & {-90 -- 90}      & -90 -- 90           \\
  \hline
 \end{tabular}
\end{table}

We next model different intrinsic source populations. Table~\ref{tab:populations} provides an overview of the required
population parameters, and the relevant result figures per population.
More information on these parameters is found in  \citet{\paperone}.

\begin{table*}
 \caption{An overview of the parameters and values used to model intrinsic FRB source populations throughout this paper. Arguments have been grouped as a subset of parameters in horizontal bands.
  Parameters include number of generated sources $n_{\rm gen}$, maximum timescale in terms of number of days $n_{\rm days}$ and whether generating a repeater population `repeaters'.
  Number density parameters $\rho$ include the number density model $n_{\rm model}$ and cosmological parameters, Hubble constant $H_0$, density parameter $\Omega_{\rm m}$, cosmological constant $\Omega_{\Lambda}$ and finally maximum redshift $z_{\text{max}}$.
  Dispersion measure (DM) components include contribution from the host DM$_{\rm host}$, from the intergalactic medium DM$_{\rm igm}$ and from the Milky Way DM$_{\rm mw}$, each with a particular model and related parameters.
  DM$_{\rm tot}$ reflects whether particular DM components are modelled or not.
  Furthermore there is the emission range $\nu_{\textrm{emission}}$, the isotropic equivalent bolometric luminosity in radio L$_{\rm bol}$, spectral index $\gamma$, intrinsic pulse width $w_{\text{int}}$ and intrinsic time stamp $t_{\text{int}}$, all with their respective modelling parameters.
  An empty space indicates a particular argument was not required for the generation of that population.
  The final row does not show arguments, but instead indicates the relevant figures per population.
  \label{tab:populations}
 }
 \centering
 \begin{tabular}{c c c c c c c}
  \hline\hline                                                                                                                                                                                  \\[-9px]
  Parameters                & Arguments          & Units                  & \pop{DM}            & \pop{rep-rate}        & \pop{rep-frac}      & \pop{complex}                                   \\[1px]
  \hline                    &                    &                        &                     &                       &                     &                                                 \\[-9px]
                            & $n_{\rm gen}$      &                        & $10^{5}$            & $10^{5}$              & $10^{5}$            & $3.6 \cdot 10^4$                                \\
                            & $n_{\rm days}$     & days                   & 4                   & 4                     & 100                 & 100                                             \\
                            & repeaters          &                        & True                & True                  & True                & True                                            \\[1px]
  \hline                    &                    &                        &                     &                       &                     &                                                 \\[-9px]
  $\rho$                    & $n_{\rm model}$    &                        & vol$_{\textrm{co}}$ & vol$_{\textrm{co}}$   & vol$_{\textrm{co}}$ & vol$_{\textrm{co}}$                             \\
                            & $\text{H}_{0}$     & km s$^{-1}$ Mpc$^{-1}$ & 67.74               & 67.74                 & 67.74               & 67.74                                           \\
                            & $\Omega_{\rm m}$   &                        & 0.3089              & 0.3089                & 0.3089              & 0.3089                                          \\
                            & $\Omega_{\Lambda}$ &                        & 0.6911              & 0.6911                & 0.6911              & 0.6911                                          \\
                            & $z_{\rm max}$      &                        & 0.01                & 2                     & 0.01                & 1                                               \\[1px]
  \hline                    &                    &                        &                     &                       &                     &                                                 \\[-9px]
  DM$_{\rm host}$           & model              &                        &                     &                       &                     & gauss                                           \\
                            & mean               & pc cm$^{-3}$           &                     &                       &                     & 100                                             \\
                            & std                & pc cm$^{-3}$           &                     &                       &                     & 200                                             \\[1px]
  \hline                    &                    &                        &                     &                       &                     &                                                 \\[-9px]
  DM$_{\rm igm}$            & model              &                        & ioka                & ioka                  & ioka                & ioka                                            \\
                            & mean               & pc cm$^{-3}$           &                     &                       &                     &                                                 \\
                            & std                & pc cm$^{-3}$           & 0                   & 0                     & 0                   & 200                                             \\
                            & slope              & pc cm$^{-3}$           & 1000                & 1000                  & 1000                & 1000                                            \\[1px]
  \hline                    &                    &                        &                     &                       &                     &                                                 \\[-9px]
  DM$_{\rm mw}$             & model              &                        &                     &                       &                     & ne2001                                          \\[1px]
  \hline                    &                    &                        &                     &                       &                     &                                                 \\[-9px]
  DM$_{\rm tot}$            & host               &                        & False               & False                 & False               & True                                            \\
                            & igm                &                        & True                & True                  & True                & True                                            \\
                            & mw                 &                        & False               & False                 & False               & True                                            \\[1px]
  \hline                    &                    &                        &                     &                       &                     &                                                 \\[-9px]
  $\nu_{\textrm{emission}}$ & low                & MHz                    & $10^7$              & $10^7$                & $10^7$              & $10^7$                                          \\
                            & high               & MHz                    & $10^9$              & $10^9$                & $10^9$              & $10^9$                                          \\[1px]
  \hline                    &                    &                        &                     &                       &                     &                                                 \\[-9px]
  L$_{\rm bol}$             & model              &                        & powerlaw            & powerlaw              & powerlaw            & powerlaw                                        \\
                            & per source         &                        & different           & different             & different           & different                                       \\
                            & low                & erg s$^{-1}$           & $10^{35}$           & $10^{40}$             & $10^{35}$           & $10^{40}$                                       \\
                            & high               & erg s$^{-1}$           & $10^{40}$           & $10^{45}$             & $10^{40}$           & $10^{45}$                                       \\
                            & power              &                        & $-1.5$              & $-1.5$                & $-1$                & 0                                               \\[1px]
  \hline                    &                    &                        &                     &                       &                     &                                                 \\[-9px]
  $\gamma$                  & model              &                        & constant            & constant              & constant            & gauss                                           \\
                            & per source         &                        &                     &                       &                     & same                                            \\
                            & mean               &                        &                     &                       &                     & $-1.4$                                          \\
                            & std                &                        &                     &                       &                     & 1                                               \\
                            & value              & ms                     & 0                   & 0                     & 0                   &                                                 \\[1px]
  \hline                    &                    &                        &                     &                       &                     &                                                 \\[-9px]
  $w_{\rm int}$             & model              &                        & constant            & constant              & constant            & lognormal                                       \\
                            & per source         &                        &                     &                       &                     & different                                       \\
                            & mean               & ms                     &                     &                       &                     & 0.1                                             \\
                            & std                & ms                     &                     &                       &                     & 1                                               \\
                            & value              & ms                     & 1                   & 1                     & 1                   &                                                 \\[1px]
  \hline                    &                    &                        &                     &                       &                     &                                                 \\[-9px]
  $t_{\rm int}$             & model              &                        & poisson             & poisson               & poisson             & poisson                                         \\
                            & rate               & day$^{-1}$             & 3                   & 3                     & 0.1                 & lognormal(9, 1)                                 \\[1px]
  \hline                    &                    &                        &                     &                       &                     &                                                 \\[-9px]
  Fig.                      &                    &                        & \ref{fig:dm_dist}   & \ref{fig:dm_rep_rate} & \ref{fig:rep_frac}  & \ref{fig:rep_frac_chime},\ref{fig:frbcat_chime} \\
 \end{tabular}
\end{table*}

\section{Results}
\label{sect:results}

\subsection{Dispersion measure distributions}
\label{sect:results:dm}
An important question in the FRB field is whether repeating and one-off FRB sources trace a single underlying population \citep[e.g.][]{petroff2019review, cordes2019review}.
One way to approach this question is to simulate a single repeating underlying FRB source population and compare the resulting observed repeating and one-off populations. \\

An initial hypothesis along these lines can be built as follows.
We assume each source produces bursts following some luminosity distribution,
where  dim  bursts outnumber bright bursts (for example: a power law with a negative index).
We only probe the part of this distribution that is above some sensitivity threshold.
The further away, the higher and more limiting the corresponding luminosity threshold becomes:
distant sources need to be far brighter to observe than sources close-by.
For repeaters this effect is stronger than for one-offs:
for repeaters at least two bursts drawn from this distribution must be seen above this threshold,
where for one-offs a single bright burst suffices.
If all FRB sources have an equal chance of emitting from a range of luminosities, one would therefore expect the observed repeating population to drop off faster with distance than the observed one-off population.
By using dispersion measure (DM) as a proxy for distance, DM distributions can be used to probe this hypothesis. \\

\begin{figure}
 \centering
 \includegraphics[width=\hsize]{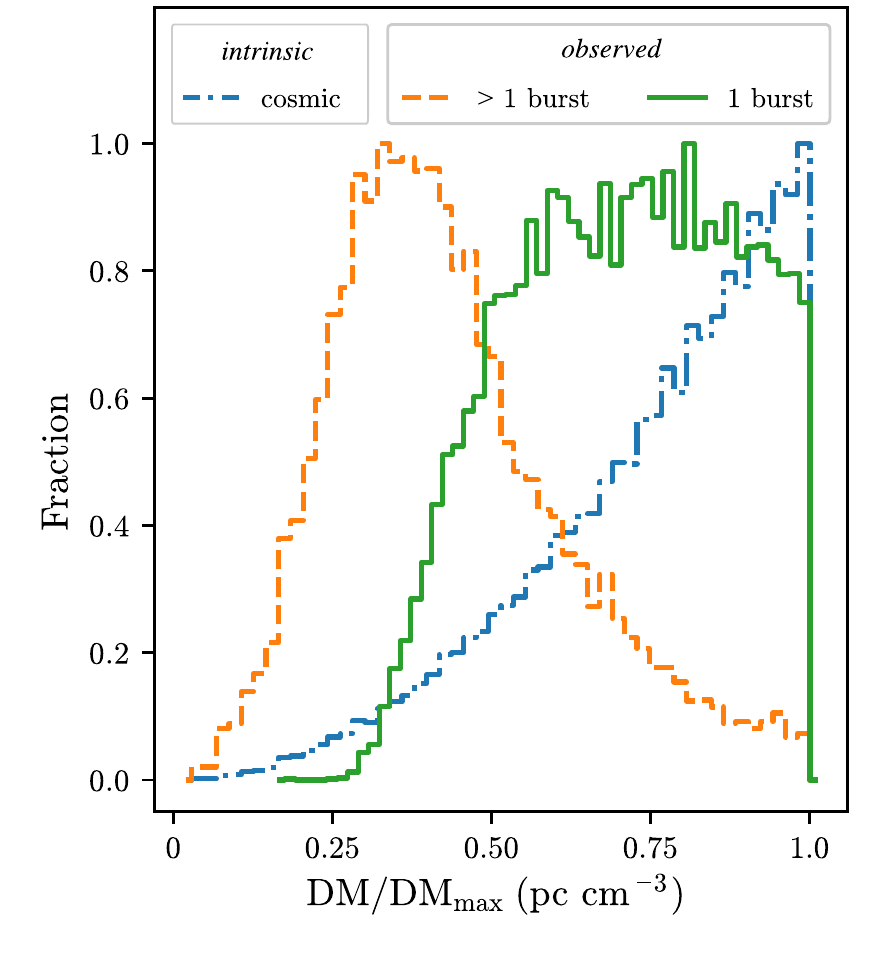}
 \caption{A comparison between the simulated intrinsic and
  simulated observed dispersion measure distributions, expressed as a fraction of maximum DM.
  Shown are the normalised simulated intrinsic (cosmic) and observed population,
  for repeaters in a Euclidean universe.
  Here the simulated observed population has been split into those seen as repeaters (>1 burst) and those seen as one-offs (1 burst).}
 \label{fig:dm_dist}
\end{figure}

We test this behaviour in Fig.~\ref{fig:dm_dist}, showing DM distributions for simulated intrinsic and simulated observed repeater populations.
The observed population has been divided into observed to be repeating sources ($>1$ burst) and single burst sources (1 burst).
For simulating this population we used parameters as given in the \pop{DM} column of Table~\ref{tab:populations}.
To avoid conflating cosmological intricacies with repeater effects, we choose to simulate the cosmic population as a Euclidean population by limiting the maximum redshift $z_{\textrm{max}}$ to 0.01.
As the absolute scale of the resulting DM distributions is not of essence, we express this scale in Fig.~\ref{fig:dm_dist} as a fraction of the total DM.
For clarity, we model the extragalactic DM contribution solely with an intergalactic component following \citet{ioka2003} in adopting $\text{DM}=1000z$ with DM in $\textrm{pc}\ \textrm{cm}^{-3}$.
Burst luminosities are drawn from a negative power law where $N(L)\propto L^{{\text{li}}}$ with ${\text{li}}$ an index of $-1.5$, in the range of $10^{35}$$-$$10^{40}$~ergs/s, and are drawn randomly per burst.
The expression for the adopted powerlaw can be converted into form of $dN(L)/dL \propto L^{1-\gamma}$ when setting ${\text{li}}=1-\gamma$ \citep[cf. the definition in][]{lu2020}.
Changing these luminosity function parameters still results in similar behaviour to that shown in Fig.~\ref{fig:dm_dist}.
For surveying this cosmic population we adopt a \survey{perfect} survey (see Table~\ref{tab:surveys}).
The \survey{perfect} survey is practically noiseless; both the noise level and luminosity boundaries are therefore mere scaling factors rather than true expectations of parameter values.
For this reason, we choose a very high S/N limit of $10^{6}$ to ensure only the high end of the flux distribution is probed.\\

Fig.~\ref{fig:dm_dist} shows our simulations predict a clear distinction between the observed DM distribution of one-offs and repeaters, despite emerging from the same cosmic population.
These distributions follow our hypothesis that the observed repeater DM distribution would be expected to tail off faster with distance than that of one-offs.
Throughout the rest of this section we refer to this expected difference as the DM discrepancy.\\

This emergence of a DM discrepancy relies chiefly on two assumptions.
Firstly that all FRB sources repeat \citep[see e.g.][]{r1,cordes2016supergiant,lyutikov2016,katz2017,metzger2019}, and secondly that the burst luminosity function is such that there are more low-energy bursts than energetic ones \citep[see e.g.][]{macquart2018papertwo,luo2020luminosity,fialkov2018}.
The results we obtain run counter to early results from CHIME/FRB \citep{fonseca2020}, which would seem to suggest that no difference is seen between the DM distribution of observed repeaters and one-offs.\\

Should a DM discrepancy remain unseen in future observations, it would lead to two possible main explanations and conclusions.\\

Firstly, a negative power law need not necessarily be an accurate representation of the luminosity function of the intrinsic source population.
Power laws are often used to approximate a wide range of physical processes, from the initial mass function \citep{1955ApJ...121..161S} to radiation from Shakura-Sunyaev thin accretion disks \citep{1973A&A....24..337S}.
While there is an abundance of FRB progenitor theories \citep{frbtheorywiki}, there is no conclusive theory on the expected emission process of an FRB.
Recent detections of FRB-like bursts from a galactic magnetar \citep[see e.g.][]{2020Natur.587...59B} may in time aid in constraining the emission mechanisms, but currently provide no prior expectation on the intrinsic luminosity function of the progenitor population.
So which luminosity functions could reduce the expected DM discrepancy?
Flatter power laws could, for instance.
The increase in number of energetic repeat bursts there leads to a higher chance of passing a S/N threshold.
This follows recent research \citep[e.g.][]{luo2020luminosity, 2020MNRAS.tmp.3335Z}, that advocate for a flatter energy, or luminosity index of respectively $-0.7$ and $-0.8$.
Simulations run with \frbpoppy for this value still however show a noticeable DM discrepancy.
Schechter functions \citep{macquart2018papertwo} also do not necessarily solve this discrepancy problem.
The asymmetrical negative trend of these functions results in the same selection effects as in negative power laws.
The DM discrepancy is avoided if the luminosity function gives repeating bursts an equal detection chance to the first detected burst.
Such functions would include for instance standard candles, or distributions that are completely flat.
Correlating observed burst luminosities with redshift estimates to FRB sources indicate this is unlikely to be the case.
Functions that are symmetric and completely visible at all distances would also explain these in principle,
but these are not necessarily in agreement with observed number counts.\\

Secondly, the lack of a DM discrepancy could arise when the source populations of one-offs and repeaters are different
in some respect.
If one-off and repeating sources occupy slightly different parts of the parameter space, selection effects will weigh
differently on both populations. This could bury the DM discrepancy.
The culprit difference between one-offs and repeaters
is not likely to be in the number density distributions
(which would follow from e.g. different progenitor populations).
In observations, repeaters and one-offs seem to trace the same DM distribution, albeit with a different normalisation \citep{fonseca2020}; such
uniformity in the DM distributions while adopting different number densities would be contrived.
The repeater population would also still be expected to show up in one-off distributions, albeit in a limited number, further complicating the situation.
One might alternatively look towards the repetition rate as a source of difference, for instance by assuming one-off
sources to intrinsically be one-offs.
The impact on detection rates resulting from e.g., a difference pulse width distributions between one-off and repeaters
might also provide a way to hide the DM discrepancy, though it is unclear how. \\

For completeness we note that the lack of an observed DM discrepancy could also be attributed to the survey. Should
CHIME/FRB be sensitive to almost all repeaters, or simply observe for long enough, the DM discrepancy would disappear,
with most repeaters being seen as repeaters.
This is unlikely to be the case, however, given the sheer number of one-offs expected to have been detected by CHIME/FRB \citep{mck20}.\\

To help constrain the origin of the lack of a DM discrepancy, we next investigate similar selection effects in repeat bursts from repeating FRB sources.

\subsection{The repeat rate dependence on DM}
\begin{figure}
 \centering
 \includegraphics[width=\hsize]{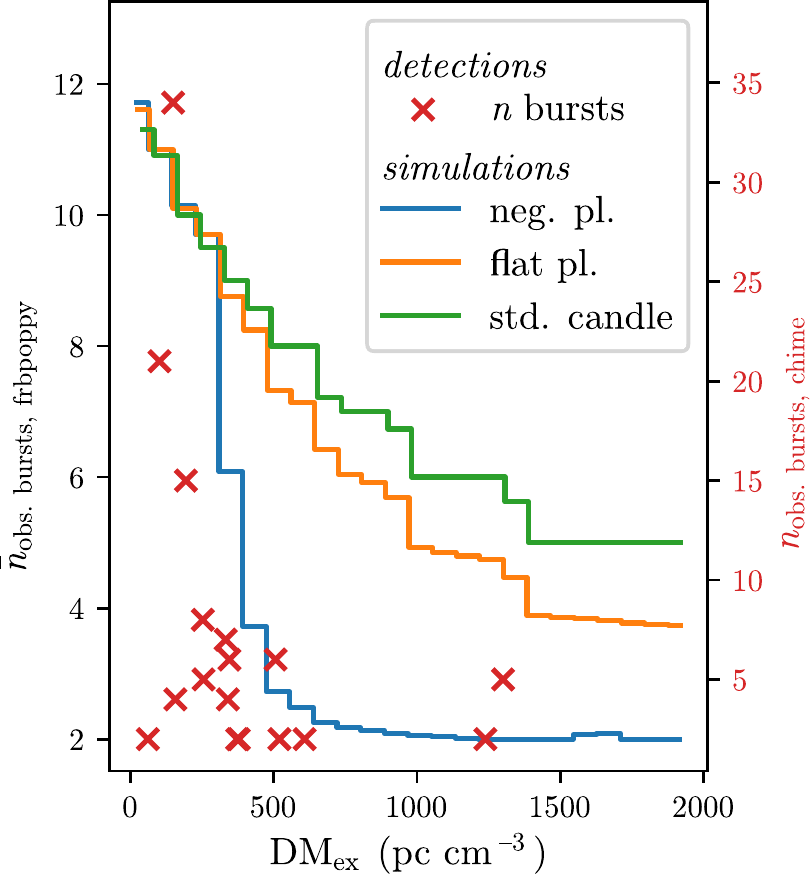}
 \caption{Observed (crosses) and simulated (lines) number of observed bursts per repeater source, as a function of extragalactic dispersion measure.
  The red crosses show the observed burst rates from CHIME repeaters (scale on right axis).
  The lines show the simulated average observed burst rates (scale on left axis showing),
  for various luminosity functions along the same extragalactic dispersion measure axis.}
 \label{fig:dm_rep_rate}
\end{figure}

If the FRB luminosity function resembles a negative power law, as in the previous section, other observed effects may also be expected.
We investigate the number of observed repeater bursts as a function of dispersion measure \citep{good2020}.
In Fig.~\ref{fig:dm_rep_rate}, the right axis marks the number of bursts per repeating source, as published in the CHIME/FRB
repeaters database\footnote{For retrieving CHIME/FRB data we make use of the pip-installable \texttt{frbcat} python package which is able to retrieve data from FRBCAT, the CHIME/FRB Repeater Database and the Transient Name Server (TNS) \citep{2020ascl.soft11011G}.} as of 2 September 2020.
For a variety of luminosity functions we compare these observations to  the simulated observed average number of bursts per source, as marked on the left axis.
Here, the negative power law population is drawn between $10^{40}$$-$$10^{45}$~ergs/s with an index of $-1.5$, the flat power law from the same range with an index of 0 and the standard candle population with bursts of $10^{42}$~ergs/s.
All other population parameters can be found in the \pop{rep-rate} column of the population list (Table~\ref{tab:populations}).
For surveying we simulate a \survey{perfect} survey.
A S/N limit of $4\cdot10^{6}$ (non-physical, as in Sect.~\ref{sect:results:dm}) provided the best visual fit in Fig.~\ref{fig:dm_rep_rate}.
Adopting a different S/N limit results in similar behaviour.\\

In this plot, we are interested in the detection-rate difference at low and high DM values.
We adopted two $y$-axes in this figure for two reasons: firstly due to the limited number of CHIME/FRB repeaters which would lead to poorly sampled bins, and secondly to allow for a relative scaling.
In our simulations we only aim to display the selection effects emerging from these luminosity functions, we have not yet aimed to reproduce the exact burst rates.
Nonetheless the behaviour of the simulated observations as shown in Fig.~\ref{fig:dm_rep_rate} still seem to suggest that selection effects due to a negative power law luminosity function better describe the observed fast drop-off of repeater burst rates with DM than a flat luminosity function.\\

A drop-off in repeater burst rates can be expected on similar grounds to the DM discrepancy presented in Fig.~\ref{fig:dm_dist}: as the distance to a source increases, the chances for a burst to fall above a S/N threshold decrease.
We therefore expect to see more bursts for close repeaters than for distant ones,
as noted in the CHIME/FRB repeater data by \citet{good2020}.
While in a Euclidean universe the average number of observed bursts over DM, or redshift, would be constant, due to the time dilation more distant sources have more bursts redshifted out of the observing time frame.
This leads to the trends seen for instance with standard candles or a flat power law, in which the average number of observed bursts drops off with distance.\\

The requirement here for a negative power law is somewhat at odds with explanation 1 from the previous section, which required a flat or symmetrical luminosity function to remove the expected DM discrepancy.
Given how the selection effects from a negative power law can describe the repeater population, and that previous studies argue for a negative asymmetrical luminosity function such as a Schechter function \citep{macquart2018papertwo,luo2020luminosity,fialkov2018}, we conclude it is likely that the FRB population can be described as a whole with a negative power law.
This leaves explanation 2 for the lack of DM discrepancy as more likely:
one-offs and repeaters subtend different parts of the intrinsic parameters space.\\

It is inviting to look to repetition rate as a potential difference in parameter space, making one-offs intrinsically one-offs or by decreasing their likelihood to repeat.
Determining the repeater fraction over time can help in establishing the veracity of such a claim.
We discuss this next.

\subsection{Repeater fraction}
The physical or environmental relationship between repeating FRB sources and seemingly one-off FRB sources is as of yet unexplained.
One line of thought is that the ostensible observed dichotomy may emerge from a single progenitor population \citep[e.g.][]{cordes2016supergiant, metzger2019, connor2020}.
Recent hints that these populations may have differing properties are however emerging, whether in pulse widths \citep{fonseca2020}, host galaxy properties \citep{2020ApJ...903..152H} or in dynamic spectra \citep{2020MNRAS.500.2525K}.
\frbpoppy can be used to probe the hypothesis of FRBs emerging from a single source population. If all FRB sources repeat with different timescales, what would the observed repeater fraction be expected to be as function of time? Could it correspond to the observed detections?\\

We start by considering how the fraction of detected sources that repeat (hereafter the repeater fraction $f_{\textrm{rep}}$) changes over time, based on two assumptions.
We assume the entire intrinsic FRB repeater source population shares a single distribution of repeat rates, such as a Poisson distribution.
We also assume a perfect survey with a S/N cut-off to limit sampling to the high end of a flux distribution.
Taking these assumptions together, one would expect the repeater fraction to asymptotically reach one - the longer you observe, the more sources you see repeat.
In the limit of infinite time, one would have seen all sources repeat.\\

A second step can be to instead introduce an intrinsic population in which all sources repeat following a Poissonian distribution, \emph{but all with a different Poissonian rate}.
To simulate this behaviour, the Poissonian rate distribution could be drawn from a normal distribution in the log space.
Here too $f_{\textrm{rep}}$ would asymptote towards one as all sources are eventually revealed to be repeaters.
Nonetheless, we expect a sharper rise, and slower tail, on the value  $f_{\textrm{rep}}$ with time, compared to the single-rate scenario.
This expectation arises from the wide range of Poisson rates - some will have a short, and others a long repetition scale.
A Weibull distribution would introduce similar behavior, albeit more extreme.
There the clustering allows for the quick detection of some repeaters. But seeing many others repeat takes far
longer,  due to the longer time intervals resulting from a Weibull distribution.\\

In an alternative scenario, the population consists of a mix of repeating and one-off sources.
How would the repetition rate differ in this case? For the repeating sources, one could expect the same behaviour as before: an asymptote towards the total fraction of repeaters.
However, as the repeater fraction reaches that asymptote it becomes increasingly likely for new detections to be one-offs.
With more and more one-offs rather than repeating sources being detected the repeater fraction will even start to show a turnover and will eventually decrease.\\

To simulate these cases, we generate a population using the parameters given in the \pop{rep-frac} column of
Table~\ref{tab:populations}, and survey this population with a \survey{perfect} survey with S/N limit of $10^4$.
We adopt population and survey parameters to reflect the most basic conditions under which these effects are still seen.
The rate parameter used as population input is varied between a delta function at 0.1 day$^{-1}$, a log-normal distribution with a rate of 0.1 day$^{-1}$ and standard deviation of 2 day$^{-1}$, and double delta function, with peaks at 0 and 0.1 day$^{-1}$, replicating a mix of one-offs and repeaters.\\

\begin{figure*}
 \centering
 \includegraphics[width=\hsize]{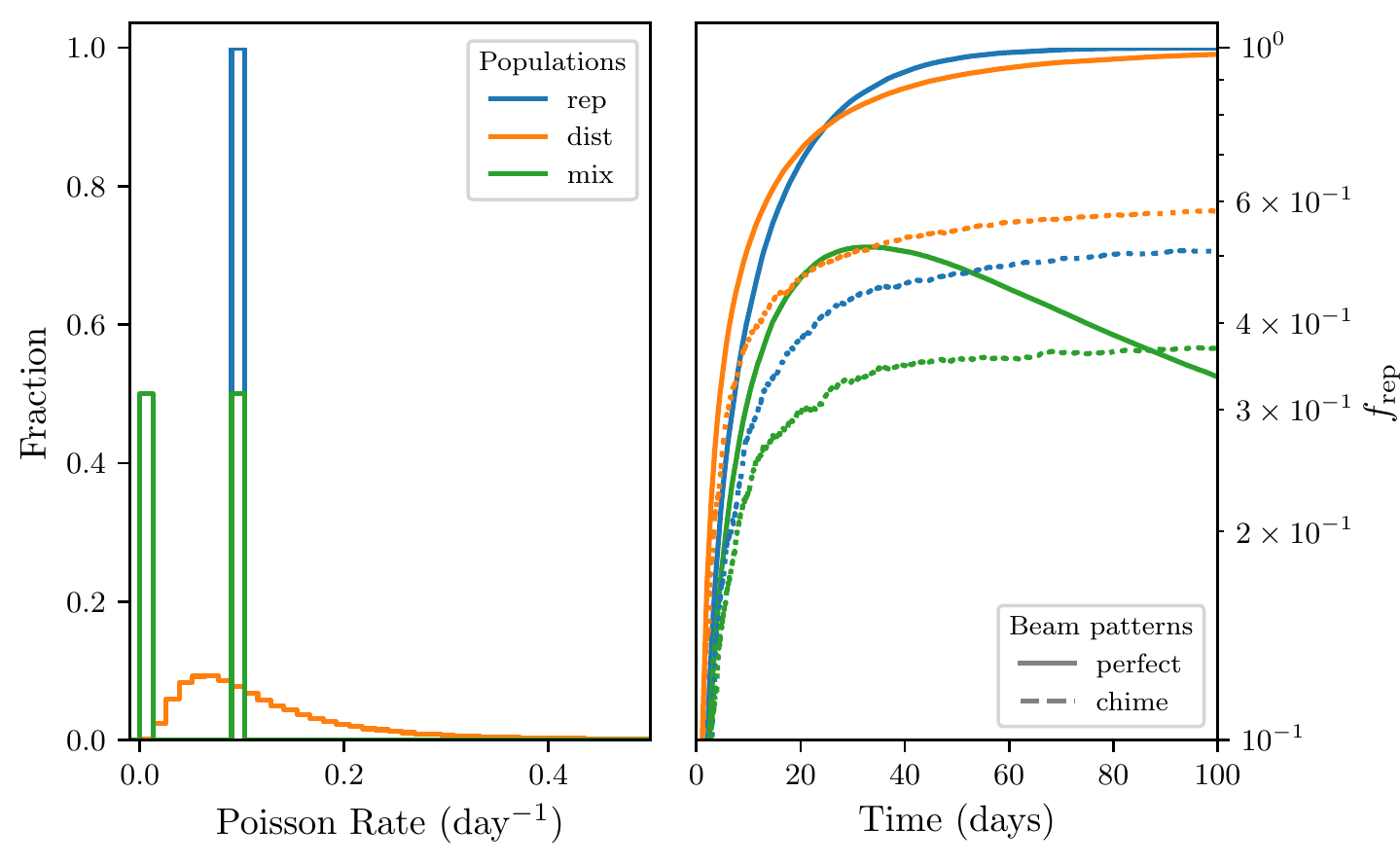}
 \caption{\emph{left:} The distributions of Poisson burst rate for various simulated intrinsic populations, including a single value (blue), a log-normal distribution (orange), and a mix of single values and one-offs (green).
  \emph{right}: Repeater fraction $f_{\text{rep}}$, defined as the number of detected repeating sources over the total number of detected sources, against time.
  The various line styles represent the detections from a perfect survey with a S/N cut-off with either a perfect beam pattern (solid), or a CHIME-like beam pattern (dotted).
 }
 \label{fig:rep_frac}
\end{figure*}

The results of these simulations can be seen in Fig.~\ref{fig:rep_frac}.
Here, the left panel shows the distributions of the mean Poisson rate given as input, while the right panel shows the change in repeater fraction over time.
For illustrative purposes we also show the results when a CHIME-like beam pattern adopted for an otherwise \survey{perfect} survey with a S/N cut-off at 1.
These latter lines show the effect of beam patterns on the observed repeater fraction.\\

The first insight that repeater fraction over time provides, lies in the expected asymptote.
This tells us the intrinsic repetition rate.
If the observed repeater fraction tends towards an asymptote at unity all FRB sources must repeat, albeit at a variety of timescales.
The speed at which the asymptote is reached contains information on the intrinsic rate distribution.
This is seen by comparing a population with a broad range of repetition rates to one with a narrow range.
For the broad ranged population, fast repeaters are detected relatively quickly, leaving the slower repeaters to be detected over a longer timescale.
The population with the same intrinsic Poissonian mean rate detects repeaters more uniformly.
This effect is still observed after applying the CHIME-like beam pattern to the simulations.
The dotted lines in Fig.~\ref{fig:rep_frac} show this effect, with distinct differences between the various intrinsic rate distributions as input, but smeared out over a longer timescale.
This smearing is why the repeater fraction of the mixed input distribution does not display a downturn when adopting a CHIME-like beam pattern; that turnover occurs beyond the timescale of this graph.
Measuring the repeater fraction over time, and by extension the intrinsic rate distribution from which it emerges, could help constrain possible rotational or orbital parameters of the repeating FRB population.
This could help rule out some of the many possible progenitor theories \citep{frbtheorywiki}.\\

The second insight comes from the value of the asymptote.
The repeater fraction show a sustained down turn over time, as seen for instance in the mixed population in Fig.~\ref{fig:rep_frac}.
This indicates one part of the FRB source population has basically completely been detected.
That is evidence for a binarity in the repeating rates of the source population.
This method would not provide any conclusive proof on a potential one-off nature of one-offs, but could only constrain the population to a maximum observed time frame.\\

Determining this trend of the repeater fraction over time observationally
will, as always, be more challenging than our \survey{perfect} survey trends in  Fig.~\ref{fig:rep_frac}.
As seen when adopting a CHIME-like beam pattern as seen in Fig.~\ref{fig:rep_frac}, selection effects muddy the trend.
Again, understanding the beam pattern of a survey to a high degree, by accurately mapping its intensity as function of
position on sky, helps in recovering a closer to intrinsic repeater fraction.
Before an asymptote or downturn is actually reached, a fit to the observed repeater fraction might already be constraining enough to determine the values of these thresholds.
This would additionally have the advantage of limiting the required observing time.
While the repeater fraction is expected to initially show a rather jagged profile due to the limited number of repeaters versus one-offs, this effect should diminish over time as more repeaters are detected.\\

A number of repeaters follow Weibull distributions \citep{oppermann2018,oostrum2020repeating}.
We investigated how such distributions might affect the repeater fraction over time.
Our simulations showed little difference compared to Poissonian rates.
Some repeaters show rapidly clustered bursts and are quickly detected as repeaters, rapidly increasing the repeater fraction.
The wait times for sources in the long tail of the Weibull distribution however
severely decrease the rate at which the asymptote is reached.

Recent results show some repeaters have period windows of burst activity \citep{chime2020periodic}.
If all repeater display such cyclic behaviour,
the repeater fraction trends would be noisier, but still display the predicted trend.\\

The results we present here are in line with those from \citet{ai2020}, who conduct similar work in investigating a repeater fraction over time.
Their simulations show a reduced complexity, which is advantageous in computational time, but lack the full range of selection effects present in \frbpoppy.
Given the strength of the selection effects in Fig.~\ref{fig:rep_frac} (cf.~Fig.~\ref{fig:rep_frac_chime}), an accurate modelling of these selection effects will be crucial in understanding the underlying source population.

\subsection{Modelling CHIME/FRB detections}
\label{sect:results:chime}
To infer the FRB progenitor population from the detected sources, we require the survey selection effects to be understood.
CHIME/FRB has detected significantly more FRBs than any other survey to date \citep{fonseca2020}.
Modelling it and its selection effects is therefore crucial for the inclusion of this dataset, the largest one available, in population synthesis with \frbpoppy.
Incorporating the CHIME/FRB detections allows insights in both the one-off population model, and the newly implemented repeater simulations.\\

As a basis for simulating an intrinsic repeating source population, we adopted the population parameters that replicate both HTRU and ASKAP-FLY one-off FRB detections \citep[see][]{2019A&A...632A.125G}.
These, and newly adopted parameters, can be found in the \pop{complex} column of Table~\ref{tab:populations}.
A number of parameters were changed with respect to the HTRU and ASKAP-FLY modelling.
We choose, for instance, to limit the intrinsic population to a maximum redshift $z_{\textrm{max}}$ of 1, a limit imposed by our compute resources.
As most FRBs have low excess DM \citep{frbcat}, suggesting low redshifts, we choose our maximum redshift as a balance between simulation size and FRB detection volume.
The adopted lower limit of the emission frequency was also increased by a single order of magnitude, to more fully sample the parameter space when adopting a negative spectral index.
We simulate each FRB source to repeat with varying luminosities and pulse widths, a choice not available when modelling one-offs.
We add modelling of the intrinsic burst time stamps. Here we adopted a lognormal distribution with a mean of 9 bursts per day, and a standard deviation of 1 burst per day.
This distribution specifically refers to the intrinsic rate distribution rather than any observed rate distributions.
To determine an optimum value for the number of sources $n_{\textrm{gen}}$, the number of days $n_{\textrm{days}}$ and the mean rate for the lognormal timestamp distribution, we ran a limited Monte Carlo simulation.
The chosen values reflect the run which best replicated the expected CHIME/FRB detection fraction of $\sim$2.5 repeating and $\sim$200 one-off sources per 100 days.
This corresponds to the expected CHIME/FRB detection rate of approximately 2 one-off sources per day \citep{chawla2017}, while 9 repeaters having been detected over just over a year \citep{fonseca2020}.
To best direct our computational resources, we only simulate FRB sources in the sky area visible to our simulated CHIME telescope, representing 67.4\% of the celestial sphere.\\

The next step is to simulate CHIME/FRB detections.
To that end, we adopted the \pop{complex} survey parameters denoted in the \survey{chime-frb} column of Table~\ref{tab:surveys}, together with the CHIME-like beam pattern described in Sect.~\ref{sec:methods:surveying}.\\

\subsubsection{Repeater fraction}
Investigating if repeating and one-off FRB sources emerge from a single progenitor population is interesting for two reasons.
First, the physics governing the burst generation, and second, the formation and evolution of the emitting sources.
If there is but one source population, its radiation mechanism would need to be capable of producing both seemingly one-off bursts and repeating bursts.
Next both one-off and repeater detections, rates, and hosts could be used to determine the progenitor population.
The question that we therefore seek to answer is: can an FRB population consisting entirely of repeaters explain the observed repeater versus one-off detection rates?\\

\begin{figure}
 \centering
 \includegraphics[width=\hsize]{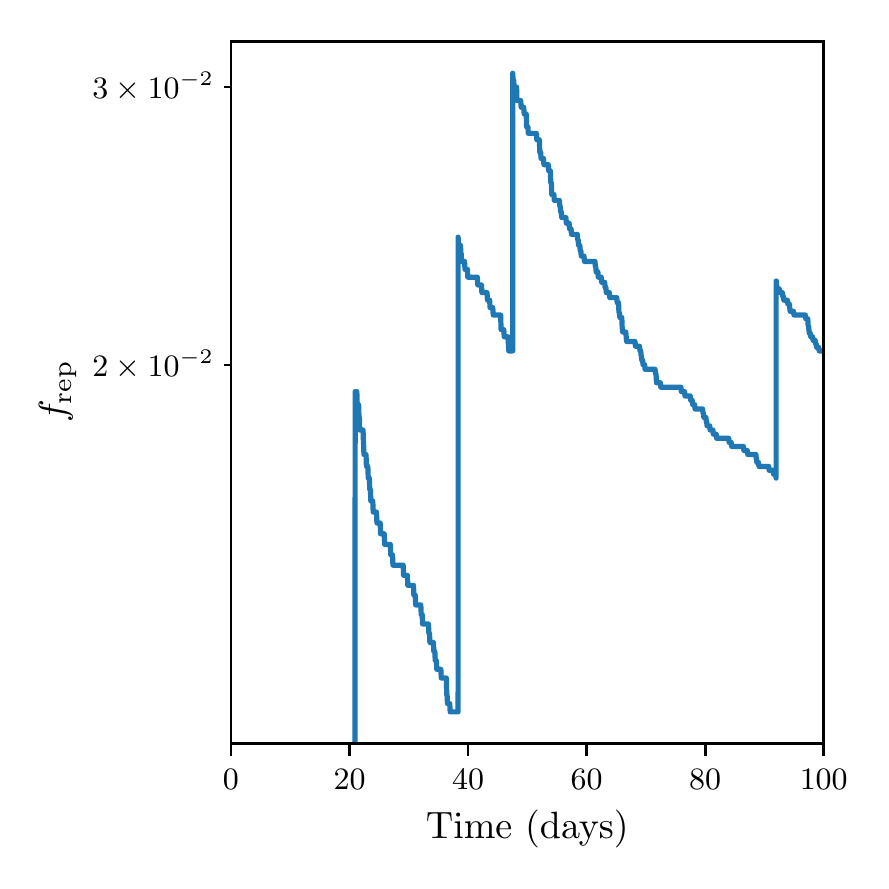}
 \caption{Repeater fraction $f_{\textrm{rep}}$, being the number of observed to be repeating sources over the total number of observed sources, against time for a full \survey{chime-frb} simulation.
 }
 \label{fig:rep_frac_chime}
\end{figure}

In Fig.~\ref{fig:rep_frac} we showed the expected repeater fraction over time for various repeater distributions.
The curves are smooth due to the high number of detections in the perfect survey: over 10$^4$ sources in 100 days.
In Fig.~\ref{fig:rep_frac_chime} we replicate this plot, but for a full CHIME/FRB simulation over 100 days using a \pop{complex} cosmic population and a \survey{chime-frb} survey.
A key difference between between the \survey{chime-frb} repeater fraction and the \survey{perfect} survey plotted in Fig.~\ref{fig:rep_frac} is the clear sawtooth effect, arising from the limited number of repeaters detected by the simulated \survey{chime-frb} over this timescale.
After 100 days, 192$_{-14}^{+15}$ one-offs and 4$_{-2}^{+3}$ repeaters are detected, close to the expected CHIME/FRB detection rate of 200 one-offs and 2 repeaters \citep{chawla2017, fonseca2020}.
The errors on the simulated values represent the corresponding $1\sigma$ Poissonian intervals.

Our \pop{complex} model is thus able to replicate the observed detection rates of both repeaters and one-offs, using an intrinsic source population consisting solely of repeaters.
Modelling the repeater fraction over time is however merely one aspect of the observed FRB population available for analysis.
Parameter distributions provide an alternative method by which the intrinsic FRB population can be probed.

\subsubsection{DM and S/N distributions}
Does the \pop{complex} model also reproduce the observed distributions of FRB parameters?
These distributions can give a handle on the progenitor population provided the selection effects are well understood.
Beyond replicating the detection rates as described above, we choose to investigate two aspects of the CHIME/FRB population --- the DM and the S/N distribution.
The DM distribution as a proxy for a distance provides a way to roughly probe the observed number density of the FRB population.
Our choice for S/N over similar parameters such as fluence, is made on the basis that it has the clearest meaning.
It convolves all observatory-based selection effects, and hence provides the cleanest comparison between survey populations \citep{2019MNRAS.483.1342J}. \\

\begin{figure}
 \centering
 \includegraphics[width=\hsize]{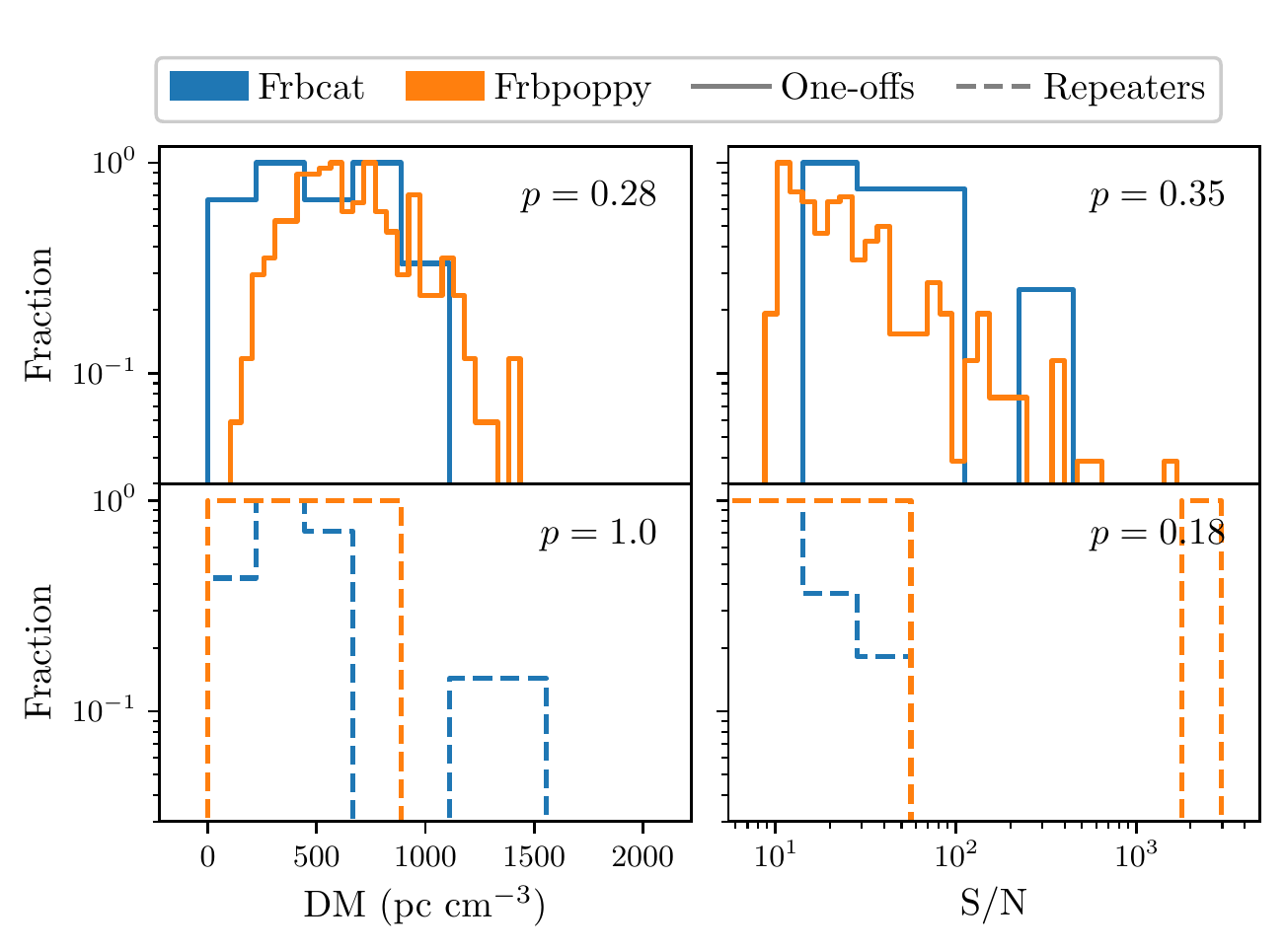}
 \caption{\emph{left column}: Real (blue) and simulated (orange) observed CHIME/FRB dispersion measure (DM) distributions for seemingly one-off sources (solid) and seen to be repeating sources (dashed).
  \emph{right column}: Same groups, but showing the observed signal to noise (S/N) distributions instead.
  In both cases, distributions have been normalised to their maximum value to allow the distributions to be compared.
  Each panel shows the $p$-values from a KS-test conducted between both shown distributions.
 }
 \label{fig:frbcat_chime}
\end{figure}

In Fig.~\ref{fig:frbcat_chime},
we show the repeater and one-off, DM and S/N distributions,
for real observed and simulated observed detections.
All distributions have been normalised to their maximum value to allow the relative shapes of the distributions to be compared.
For the repeaters, we plot the average DM value and the S/N of the first detected burst per source.
We use only the first burst to avoid a bias arising from a single source saturating distributions with a high number of bursts.
A Kolmogorov–Smirnov (KS) test is used to compare each set of distributions of which the result is given in the top right of each panel.
The real observed distributions were obtained  from \texttt{frbcat} and the CHIME/FRB repeater database as of 2 September 2020.
The simulated observed distributions use the \pop{complex} model.
Fig.~\ref{fig:frbcat_chime} shows a run from the small Monte Carlo simulation (Sect. \ref{sect:results:chime}) with
the high KS-test output.
The $p$ values are all above 0.05.
Given the limited number of trials in the simulation, these $p$ values indicate the observations and simulations are consistent with being drawn from the same distribution.
While these results are clearly based on very small numbers, they do indicate that the \pop{complex} model can explain the observed CHIME/FRB populations to a reasonable degree.\\

The simulated and real observed DM distributions for both one-off and repeating FRB sources are seen in the left column of Fig.~\ref{fig:frbcat_chime}.
Although the limited number of detected repeaters necessarily makes comparisons challenging,
the KS-tests for these parameters indicate an encouraging match between observations and simulations for our
\pop{complex} model.
In our simulations, the one-off and repeater populations span similar parts of the DM space; perhaps contrary to expectations on basis of Fig.~\ref{fig:dm_dist}.
The reason is that the \pop{complex} model underlying these simulated populations uses a flat luminosity index,
which was shown to be able to replicated observed HTRU and ASKAP-FLY one-off detections,
while the results given in Fig.~\ref{fig:dm_dist} explored the impact of a  negative index.
The lack of a DM discrepancy corresponds to that seen in the CHIME/FRB data, in which both one-offs and repeaters are observed to follow the same distribution \citep{fonseca2020}.\\

The S/N distributions for the simulated and observed repeater and one-off populations can be found in the right column of Fig.~\ref{fig:frbcat_chime}.
The simulations fit these distributions, too.
The slopes for the one-off distributions are similar.
There is a noticeable difference at low S/Ns, where \frbpoppy expects more low S/N events than observed.
We conclude CHIME becomes incomplete below S/N$\simeq$15.
Indeed, comparing only detections above a S/N limit of 15 gives a much improved fit, with a $p$-value of 0.98 for the one-offs.
Potential explanations for the incompleteness are that the CHIME beam pattern is less sensitive than our simulated
CHIME-like beam pattern, or that e.g., the RFI mitigation techniques adopted by CHIME/FRB block real, low S/N events \citep{chimeoverview}.
Especially for one-offs it can be challenging to determine if a candidate is real, and lower S/N detections might therefore be disregarded out of caution.
In repeaters however, the same low-S/N candidate would be marked real if prior bursts were detected at the same DM and location.
For this reason it can be important to compare possible selection effects in the detection pipeline of various surveys with for instance the benchmarking test set up for FRB detection pipelines \citep{frbolympics}, similar to prior work with pulsar pipelines \citep{2015ApJ...812...81L}.
The simulated repeater distributions similarly show a difference, with repeaters showing up at high S/N.\\

Reasons for caution in interpreting these fits is first, the low number of repeaters, just four, over this timescale;
and second, the short  simulation span of 100 days, while the real CHIME/FRB observations span a multiple thereof.
The single high S/N event showing up in the simulated repeater distribution (Fig.~\ref{fig:frbcat_chime}) is a curious appearance, and on basis of prior simulations we believe it could be indicative of a slope more in line with that of the one-offs.
That is in contrast with the observed CHIME distributions, where repeaters seem to show a steeper S/N distribution than one-offs.
Including more, newly published CHIME detections will help with investigating this observed discrepancy, and determining the origin thereof.\\

An interesting statistical distinction between repeaters and one-off events is emerging in the CHIME data set.
One-off FRBs appear to have narrower pulse widths than sources that have been detected twice or more \citep{fonseca2020}.
This effect may be due to an intrinsic difference between repeaters and non-repeaters, or due to an observational bias, as suggested by \citet{connor2020}.
The effect does not appear in \frbpoppy.
This is unsurprising, because we neither model the FRB population as two separate source classes with different average widths nor include beaming effects.\\

The simulations of the detection rates seen in Fig.~\ref{fig:rep_frac_chime}, and the DM and S/N distributions seen in Fig.~\ref{fig:frbcat_chime}  match the observed CHIME/FRB population.
As these \pop{complex} population parameters also resulted in good fits to the observed one-off populations by HTRU and ASKAP \citep[see][]{\paperone}, they provide a solid basis from which the intrinsic FRB parameter space can further be explored.
These fits additionally provide a good indication that a purely repeating population could describe the observed FRB populations.
If so, observations focussing on one-off theories such as double neutron star mergers \citep{totani2013}, double white dwarf mergers \citep{kashiyama2013} or similar cataclysmic models could be dropped in favour of following expected observational signatures from repeating models such as from young magnetars \citep{metzger2019}, flares from magnetar wind nebulae \citep{beloborodov2017} or other models \citep[see][]{frbtheorywiki}.\\

Further exploration of parameter spaces with \frbpoppy are part of a subsequent investigation \citep{gardenier21};
further interpretation of the resulting  population parameters will be carried out when CHIME FRBs are published.\\

\subsection{Opportunities, uses and further work}
Our results demonstrate the value of FRB population synthesis, also for repeating sources.
\frbpoppy is open source by nature to encourage such use of FRB population synthesis.
It can power research avenues ranging from  simulations of the effect of different input distributions, to comparison studies of detections across various surveys.
One clear result often cropping up in our simulations is the importance of the beam pattern.
Shifting from extensive observations of single sources to probing the full FRB population will require team derived and publishing accurate beam patterns.
Beam pattern mapping has mostly been a focus of the imaging domain, yet understanding the effects of beam pattern on FRB detections will allow for far better probing of the intrinsic parameter space of the FRB source population.
This would help in collating observations from multiple surveys to form a single, coherent picture of the FRB population.

\section{Conclusions}
\label{sect:conclusion}
We wanted to investigate if one-offs and repeaters can emerge from the same intrinsic source population, and if selection effects explain the observed differences.
We thus implemented repeating FRB sources in \frbpoppy, an open source FRB population synthesis package in Python.
We conclude that:

\begin{enumerate}
 \item Our simulations can reproduce current multi-survey observational data by synthesising a population of only
   repeating FRBs, provided they have a wide distribution of repetition rates,
 \item The luminosity function of FRBs can significantly impact the observed DM distribution of repeaters versus one-off detections (i.e.~apparent non-repeaters).
       Should the DM distributions of repeaters and one-offs remain in agreement, as hinted at in CHIME data, and evidence continue to point towards an intrinsic luminosity function described by a negative power law with more dim bursts than energetic ones, it could potentially suggest the presence of an intrinsic difference between repeating and one-off sources.
 \item Within the observed repeater population, frequent repeaters tend to be closer and have smaller DMs.
       This effect was noticed in CHIME data by \citet{good2020}, and we use \frbpoppy to explain the inverse relationship between DM and repetition rate.
       The relationship is a consequence of point 2 and indicates that the luminosity function of repeating FRBs is given by a negative power law with more dim bursts than energetic ones.
 \item FRB surveys can use the observed repeater fraction over time to determine whether there is any binarity in the intrinsic repetition rate of the FRB source population.
       \frbpoppy is the ideal tool for such an exercise because it can account for instrumental selection effects that are difficult to model analytically.
\end{enumerate}

Overall we thus find that the observed FRB sky can be
explained by a single population of repeating FRBs that is uniform in its major characteristics,
but where the repeat rate correlates with other, more minor, behavioral or physical traits.

\begin{acknowledgements}
 We thank the participants of FRB2020 for the fascinating conference, and especially Deborah Good for the illuminating discussion.\\
 \ \\
 The research leading to these results has received
 funding from the European Research Council under the European Union's Seventh Framework Programme (FP/2007-2013) / ERC
 Grant Agreement n. 617199 (`ALERT'); from Vici research programme `ARGO' with project number
 639.043.815, financed by the Netherlands Organisation for Scientific Research (NWO); and from
 the Netherlands Research School for Astronomy (NOVA4-ARTS).
 EP further acknowledges funding from an NWO Veni Fellowship.\\
 \ \\
 We acknowledge use of the CHIME/FRB Public Database, provided at \url{https://www.chime-frb.ca/} by the CHIME/FRB Collaboration. We additionally acknowledge the use of the FRB catalogue `frbcat' \citep{frbcat}, available at \url{www.frbcat.org}, and the use of NASA’s Astrophysics Data System Bibliographic Services.\\
 \ \\
 This research has made use of \package{python3}{python} with \package{numpy}{numpy}, \package{scipy}{scipy}, \package{astropy}{astropy}, \package{pandas}{pandas}, \package{matplotlib}{matplotlib}, \package{bokeh}{bokeh}, \package{requests}{requests}, \package{sqlalchemy}{sqlalchemy}, \package{tqdm}{tqdm}, \package{joblib}{joblib}, \package{frbpoppy}{\paperone} and \package{frbcat}{2020ascl.soft11011G}.
\end{acknowledgements}

\bibliographystyle{aa}
\bibliography{bibliography}

\begin{appendix}
 \section{Tracking celestial objects}
 \label{sec:appendix}
 Determining the path of a celestial object through a beam pattern is a non-trivial challenge.
 To simulate the surveying of a repeating FRB population, \frbpoppy incorporates functions to calculate the position of objects within a beam pattern.
 In \frbpoppy we approach this challenge by transforming source coordinates to the coordinate system relative to the beam pattern.
 These transformations differ depending on the type of telescope mount involved.\\

 In \frbpoppy we choose to model beam patterns in 2D matrices, yet we generate source coordinates in a (3D) equatorial coordinate system.
 Here we use 3D to refer to a coordinate system such as right ascension and declination, which per definition describe angles on a unit sphere in 3D space.
 Determining the position of celestial object in a beam pattern therefore requires a mapping to be made from three to two dimensions.
 Adopting a gnomonic projection for this transformation allows a beam pattern to be expressed in units of angular offset (degrees) relative to a central pointing.
 We follow \citet{snyder1987} in expressing the gnomonic projection as
 \begin{eqnarray}
  \label{eq:gnomonic-start}
  \cos \Delta x & = & \cos x_{\rm ref} \cos x_{\rm obj} + \sin x_{\rm ref} \sin x_{\rm obj}\\
  \sin \Delta x & = & \cos x_{\rm ref} \sin x_{\rm obj} - \sin x_{\rm ref} \cos x_{\rm obj}\\
  \cos c & = & \sin y_{\rm ref} \sin y_{\rm obj} + \cos y_{\rm ref} \cos y_{\rm obj} \cos \Delta x\\
  \Delta x & = & \frac{\cos y_{\rm obj} \sin \Delta x}{\cos c}\\
  \Delta y & = & \frac{\cos y_{\rm ref} \sin y_{\rm obj} - \sin y_{\rm ref} \cos y_{\rm obj} \cos \Delta x}{\cos c}
  \label{eq:gnomonic-end}
 \end{eqnarray}
 with $\Delta x$/$\Delta y$ the orthogonal offset in 2D, $x_{\rm ref}$/$y_{\rm ref}$ the 3D reference (or pointing) angular coordinates, and $x_{\rm obj}$/$y_{\rm obj}$ the 3D object angular coordinates. Note that these equations are only valid when $\cos c >= 0$, being undefined when $\cos c < 0$. This limit represents pointings on the sky beyond the observable horizon of a telescope located on a sphere.\\

 For observatories with equatorial mounts, celestial objects remain in a constant position with respect to the beam pattern of a single pointing of a single dish telescope. As such, the right ascension $\alpha$ and declination $\delta$ can be adopted directly as respectively $x$ and $y$ in Eq.~\ref{eq:gnomonic-start}--\ref{eq:gnomonic-end}. This avoids any additional transformations of the reference and object coordinates.\\

 Azimuthally mounted telescope however, do not retain a constant angle with respect to the North Pole, leading objects to wander through a beam pattern over the course of a single pointing. In \frbpoppy we model detections by such telescopes by shifting source pointings from an equatorial to an azimuthal coordinate system. We assume a survey to start at a random point of time in this century, and for both the relative and object coordinates calculate the local hour angle:
 \begin{equation}
  \text{LHA} = \text{LST} - \alpha
  \label{eq:lha}
 \end{equation}
 with local hour angle LHA, local sidereal time LST and right ascension $\alpha$. Taking this together with the declination $\delta$ and the latitude of a telescope $\lambda$ allows the altitude Alt and azimuth Az to be calculated:
 \begin{eqnarray}
  \label{eq:alt}
  \text{Alt} &=& \arcsin \left( \sin \delta \sin \lambda + \cos \delta \cos \lambda \cos \text{LHA} \right)\\
  \text{Az} &=& \arccos \left( \frac{\sin \delta-\sin \text{Alt} \sin \lambda}{\cos \text{Alt} \cos \lambda} \right)
  \label{eq:az}
 \end{eqnarray}
 For $\text{LHA}>0$ an East-West correction has to be applied in the form of $\text{Az}=360-\text{Az}$.
 The resulting azimuth and altitude of both the object and the reference point can subsequently be used as $x$ and $y$ in Eq.~\ref{eq:gnomonic-start}--\ref{eq:gnomonic-end}.\\

 Transit observatories can be modelled in a similar fashion to azimuthally mounted telescopes. By fixing the reference coordinate to the zenith, source coordinates can be transformed to the azimuthal coordinate system using Eq.~\ref{eq:lha}--\ref{eq:az} before plugging into Eq.~\ref{eq:gnomonic-start}--\ref{eq:gnomonic-end}.
\end{appendix}

\typeout{get arXiv to do 4 passes: Label(s) may have changed. Rerun}
\end{document}